\documentclass{PoS}

\usepackage{amsmath}
\usepackage{amssymb}
\usepackage{graphicx}
\usepackage{wrapfig}

\newcommand{\de}{\delta}

\title{Probing New Intra-Atomic Force with Isotope Shifts: \\ A Neat Thing to Do}

\ShortTitle{Probing new intra-atomic force with isotope shifts}

\author{\speaker{Yasuhiro YAMAMOTO}\thanks{The talk is based on a part of Ref.~\cite{Mikami:2017ynz}}\\
        Department of Physics and IPAP, Yonsei University, Seoul 03722 Republic of Korea\\
        E-mail: \email{yamayasu@yonsei.ac.kr}}

\abstract{
 We discuss a new method to search for a new very weakly interacting light boson with extremely precise atomic spectroscopy, namely, the atomic clock.
 The contribution of the new physics may appear as the violation of a linear relation of the isotope shift.
 We evaluated this effect with some simple assumptions.
 Since the results still have disagreements with works by the other group, we briefly mention some points which should be improved in the future.
}

\FullConference{ICHEP 2018, International Conference on High Energy Physics\\
		4-11 July 2018\\
		Seoul, Republic of Korea}

\begin{document}

\section{Introduction and results}
\begin{wrapfigure}[31]{r}{19em}
%\begin{figure}[b]
%\centering
\vspace*{-\intextsep}%
\includegraphics[width=18em,clip]{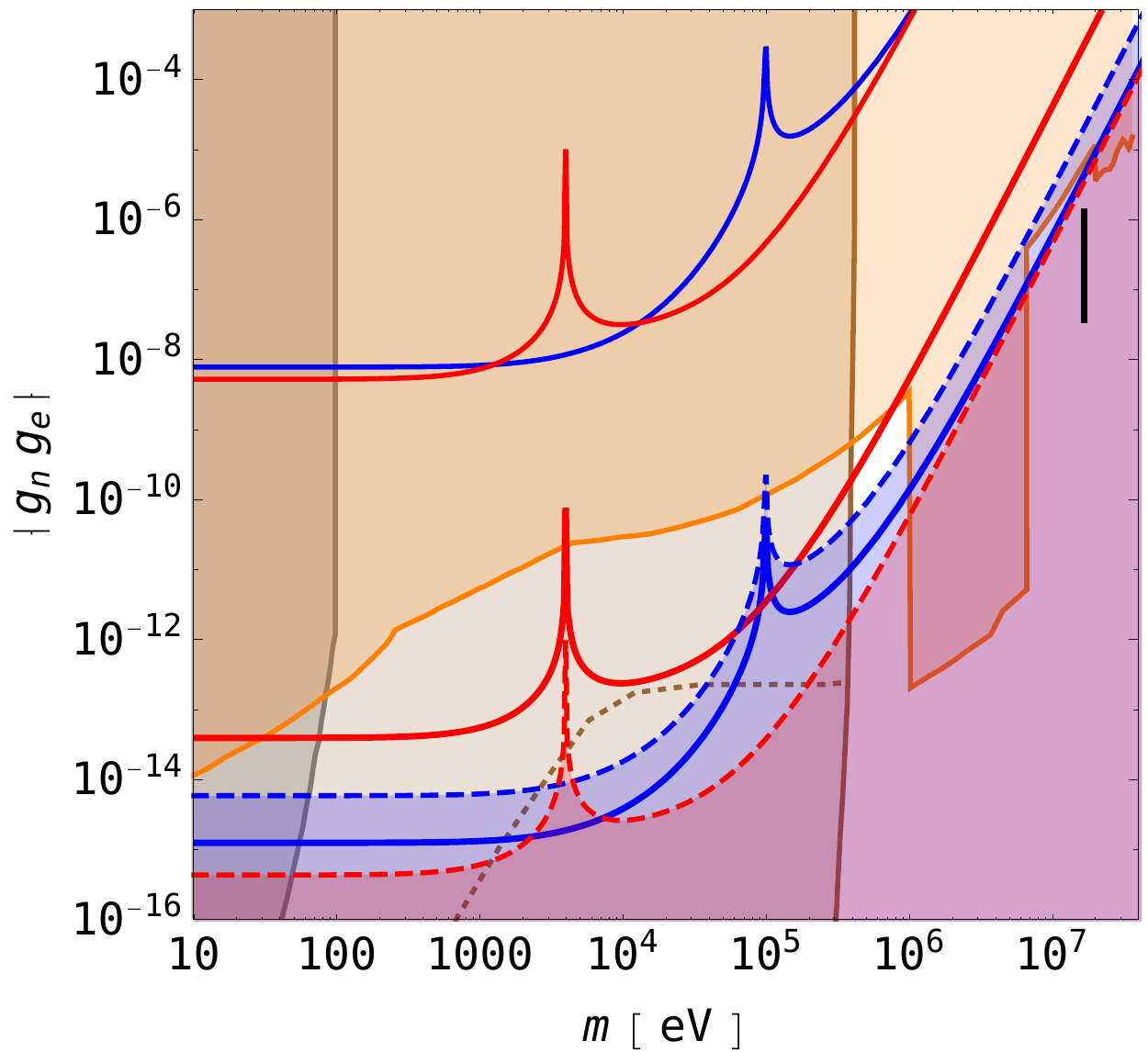}
\caption{
  The isotope shift and other experimental constraints on the mass and the coupling of the light vector.
	The red/blue lines and region are used for the bounds with the isotope shifts by Ca$^+$/Yb$^+$.
	The upper solid lines are the current experimental bounds, and the lower ones are the future prospects with the error of 1 Hz.
	The non-linearities by the NLO field shifts appear in the shaded regions below the dashed lines.
	The upper shaded orange regions indicate the constraints obtained with the bounds on the couplings of electron and neutron by low energy experiments, see Ref.~\cite{Mikami:2017ynz} for the details.
	The shaded brown regions below about 500 keV is constrained by the stellar cooling bounds given by Ref.~\cite{An:2013yfc}.
	As described in Ref.~\cite{Redondo:2008aa}, the stellar bounds have uncertainty above the brown dotted line.
	The shaded gray regions below about 100 eV are restricted by the fifth force experiment~\cite{Ederth:2000zz,Fischbach:2001ry}.
The black line in the right panel stands for the region indicated by the Atomki anomaly~\cite{Krasznahorkay:2015iga}.
}
\label{FigVector}
\end{wrapfigure}
%\end{figure}

In low energy experiments, the very weakly interacting light particles can give us the signals similar to that given by the heavy particle with $O(1)$ couplings.
Therefore, in order to explain some low energy anomalies, those light particles attract our attention as well as the 10 TeV physics investigated by LHC.
As a possible new strategy to study the very weak new force, we use the atomic clock experiments which give us extremely precise results on the atomic spectrum.

The isotopes basically posses the same electromagnetic property.
Tiny differences come from the states of the nuclei.
At the consequence, the spectra are slightly different i.e. so-called the isotope shift.
The leading contribution of the shifts preserves a linear relation~\cite{RefKing}, the King linearity.
However, new physics effects do not have to do.
Therefore, the violation of the linear relation can be the probe of the new physics.
This possibility was started to discuss by Ref.~\cite{1601.05087} and their follow-up papers.
They roughly estimated the new constraints with their qualitative intuition and numerical tools.
We have quantitatively studied the same phenomena with some assumptions to simplify the calculation.
A result of our study is shown in Fig.~\ref{FigVector}.

In the rest of the article, we briefly introduce what happens in the isotope shift and the non-linearity following Ref.~\cite{Mikami:2017ynz}.
Compared with the other works, we simultaneously point out what we should improve in the future.

\section{A neat thing to do}

The isotope shift $\de\nu$ for a transition can be given as
\begin{align}
 \de\nu = G \de\mu +F\de\langle r^2 \rangle +X,
\label{EqShift}
\end{align}
where the first two terms are the leading contributions of the isotope shift, so called the mass and the field shift, and $X$ is the others including the NLO contributions and new physics effects.
The difference of the reduced masses and the mean square radii are given by $\de\mu$ and $\de\langle r^2 \rangle$, respectively.
The coefficients $G$ and $F$ depend on the wave function of each state in transitions.
Using two different transitions, we obtain
\begin{align}
 \frac{\de\nu_2}{\de\mu} =\frac{F_2}{F_1} \frac{\de\nu_1}{\de\mu} +G_2 -\frac{F_2}{F_1}{G_1} +\frac{1}{\de\mu} \left( X_2 -\frac{F_2}{F_1} X_1 \right).
\end{align}
The plot of $\de\nu / \de\mu$ for several isotopes preserve the linear relation, unless $X$ is not included.

We introduce the extra contributions in the isotope shift which are written as
\begin{align}
 \Theta = \int dr\, r^2 \left( R_{N'}(r)^2 -R_N(r)^2 \right) V_\Theta (r),
 \label{EqInt}
\end{align}
where $R_N$ is the radial wave function with the set of the quantum numbers $N$.
If $V_\Theta$ is the modification of the Coulomb potential for different isotopes, $\Theta$ is the field shift.
Evaluating it with the expansion of the wave functions, we obtain the Seltzer moment expansion~\cite{PhRva.188.1916}.
The leading contribution is $F\de\langle r^2 \rangle$, and the rest is $X$ in Eq.~\eqref{EqShift}.
The NLO field shift is shown in Fig.~\ref{FigVector}~\footnotemark.
\footnotetext{
 This is a different NLO effect discussed in Ref.~\cite{1709.00600}.
 The Standard Model effects have been originally studied very well in Ref.~\cite{RefFs}.
}
If $V_\Theta$ is the Yukawa potential induced by the new force, $\Theta$ is the particle shift.
We see some behaviours of the particle shift below.

If the mass of the new boson is light enough, i.e., less than about 10 keV according to Fig.~\ref{FigVector}, Eq.~\eqref{EqInt} is similar to the expectation value of the Coulomb potential.
In this mass region, the sensitivity to the coupling becomes flat because the value of the integration $\Theta$ saturates.
We have estimated the wave functions as the one-electron problem in the electron cloud.
The cloud is calculated by the Thomas-Fermi model.
This treatment is much simpler than the other calculations given by Ref.~\cite{1704.05068}.
However, the obtained results are numerically similar.

We discuss heavy mediator which is heavier than the nuclear scale, namely, about several fm.
In this case, we can evaluate the integral with the expansion of the wave functions such as the field shift.
The expansion can be written with a notation as,
\begin{align}
 \left( R_{N'}(r)^2 -R_N(r)^2 \right) = \xi_0 +\xi_1 r +\xi_2 r^2 +\cdots.
\end{align}
The integration with the Yukawa potential and the term of $r^n$ gives us the new physics contribution proportional to $m^{-2-n}$, i.e., the leading contribution is proportional to $m^{-2}$.
However, this term gives us term proportional to $\de\langle r^2 \rangle$ as the leading field shift.
Then, its effect is absorbed in the shift of $F$, that is, it does not violate the linearity.
\footnotetext{
 Any Higgs contribution cannot be seen in these experiments.
}
The next leading contribution seems to be the term of $m^{-3}$ as Ref.~\cite{1704.05068} has been stated.
However, the wave function becomes flat at the origin if we consider the volume effect of the nuclear charge distribution.
Therefore, the leading effect to violate the linearity appears as the term proportional to $m^{-4}$.
This means that the sensitivity to the mediator heavier than about tens MeV rapidly gets worse.

Finally, we discuss the region between the above two.
In this region, the wave function can be estimated with the potential generated by the point charge.
This means that the $m^{-3}$ term given by an $s$-state seems to be the leading contribution.
In the atomic-clock experiments, the same $s$-states are usually included in the transitions to draw the linearity plot.
In this case, the contribution trivially is cancelled.
This disappearance of the $s$-state contribution also happens for the case that just one $s$-state is included in the transitions.
The leading contribution by $p$-states is the term of $m^{-4}$ which is shown in Fig.~\ref{FigVector}.

In the discussions above, we have not considered the relativistic effects.
Since the typical elements employed in the experiments are heavy atoms such as Ytterbium, the relativistic effect can be large and change the qualitative properties of the discussions.
The discussion in Ref.~\cite{1709.00600} includes some details of the relativistic behaviour.
Since the relativistic effect makes the incline of the wave function steeper, they have concluded that the sensitivity to the heavier mass region become better such as $m^{-1}$\footnotemark.
\footnotetext{
 In the v1 of the arXiv article, the scaling property of the sensitivity was similar to our non-relativistic result.
}
The results itself cannot be checked with our non-relativistic calculation.
However, it is worth considering the above cancellation of the $s$-state.

\section{Conclusion}

We discussed the possibility to probe the very weakly interacting light boson with the atomic spectroscopy.
Their constraints to the electron-neutron interaction can be stronger than the other terrestrial experiments in the near future.
In this article, we focus on the behaviour of the particle shift.
Our results still have some inconsistency with the other's ones.
In order to obtain more precise results on the search potential and clarify the structure of this phenomena, we need to investigate the details of relevant atomic physics.

\end{document}